\documentclass[aps,prl,twocolumn,superscriptaddress,showpacs]{revtex4}
\usepackage{amsmath,amssymb,graphicx,color,textcomp}

\begin{document}

\title{Spontaneous Pattern Formation in a Polariton Condensate}
\author{F.~Manni}
\email{francesco.manni@epfl.ch}
\author{K.~G.~Lagoudakis}
\affiliation{Institute of Condensed Matter Physics, Ecole Polytechnique F\'{e}d%
\'{e}rale de Lausanne (EPFL), CH-1015, Lausanne, Switzerland}
\author{T.~C.~H.~Liew}
\affiliation{Institute of Theoretical Physics, Ecole Polytechnique F\'{e}d\'{e}rale de
Lausanne (EPFL), CH-1015 Lausanne, Switzerland}
\author{R.~Andr\'e}%
\affiliation{Institut N\'eel, CNRS, Grenoble, France}
\author{B. Deveaud-Pl\'edran}
\affiliation{Institute of Condensed Matter Physics, Ecole Polytechnique F\'{e}d%
\'{e}rale de Lausanne (EPFL), CH-1015, Lausanne, Switzerland}

\begin{abstract}
Polariton condensation can be regarded as a self-organization phenomenon, where phase ordering is established among particles in the system. In such condensed systems, further ordering can possibly occur in the particle density distribution, under particular experimental conditions. In this work we report on spontaneous pattern formation in a polariton condensate under non-resonant optical pumping. The slightly elliptical ring-shaped excitation laser we employ is such to force condensation to occur in a single-energy state with periodic boundary conditions, giving rise to a multi-lobe standing wave patterned state.
\end{abstract}

\date{\today}
\maketitle



Self-organization and pattern formation phenomena are among the most cross-sectional topics in science, as they range from the spontaneous folding of proteins in biology \cite{karsenti_self_2008} and self-assembly of molecules \cite{whitesides_molecular_1991} to liquid crystal ordering in chemistry \cite{lehn_self_2002,ungar_giant_2003}. The spontaneous symmetry breaking that accompanies polariton condensation \cite{balili_bose_2007,deng_spatial_2007,kasprzak_condensation_2006,wertz_spontaneous_2010,lai_coherent_2007} and superfluidity \cite{amo_superfluidity_2009,amo_collective_2009} can be interpreted as the manifestation of a self-organization process of particles in the system, where phase ordering is established at macroscopic scales. In such systems, under particular conditions, additional ordering may possibly occur in the density distribution, giving rise to spatial patterns. In this work we report on the spontaneous spatial pattern formation in an exciton-polariton condensate, under non-resonant optical excitation with a slightly elliptical ring-shaped monomode laser beam. The very shape of the excitation ring forces condensation to take place in a quasi one-dimensional single energy state with periodic boundary conditions. The cylindrical symmetry of the system is broken by the ellipticity of the excitation intensity profile. This causes polaritons to condense in an elliptical multi-lobe patterned single-energy state.

Polaritons, half-light half-matter quasiparticles, represent the eigenmodes of strongly interacting light and matter, which can be achieved in planar semiconductor microcavities endowed with quantum wells, in the strong coupling regime~\cite{weisbuch_polariton_1992}. Spatial localization of polaritons can be achieved on the micrometer scale due to their small effective mass~\cite{kasprzak_condensation_2006}, coming from their photonic component. A successful approach, in order to study the polariton phenomenology in lower dimensionality, consists in the localization of polaritons via the confinement of the photonic component through nanostructuring of the planar two-dimensional microcavity, leading to both 1D and 0D polariton states, as in~\cite{wertz_spontaneous_2010,cerna_coherent_2009}. In these works the confinement of the polariton wave function comes from the real potential created by the structure realized in the samples. Other studies have rather focused on polariton localization effects and spatial inhomogeneous condensation due to the presence of photonic disorder, a feature naturally arising in semiconductor microcavities as a result of the growth processes of the samples. Disorder and multi-mode condensation in spatially overlapping modes has been extensively studied~\cite{krizhanovskii_coexisting_2009}, whilst a laser-like gain-induced mechanism for polariton localization because of the excitation spot has also been reported~\cite{roumpos_gain_2010} in a disordered potential.

In this letter we demonstrate an all-optical spontaneous patterning of the density of a polariton condensate within the ring geometry of the non-resonant excitation laser. The laser acts, through a gain-loss mechanism, as an optical trap with periodic boundary conditions. Under these experimental conditions, we report on the spontaneous formation of a multi-lobe ring patterned polariton condensate, with similar features to the ones seen in polariton microwires~\cite{wertz_spontaneous_2010}. We non-resonantly create the polariton condensate and, thus, we do not imprint any phase distribution with the pump laser. The observed phenomenology and the pattern formation mechanism are reproduced through a theoretical model based on the generalized Gross-Pitaevskii equation (GPE)~\cite{wouters_spatial_2008}.

\begin{figure}[tb]
\includegraphics[width=0.48\textwidth]{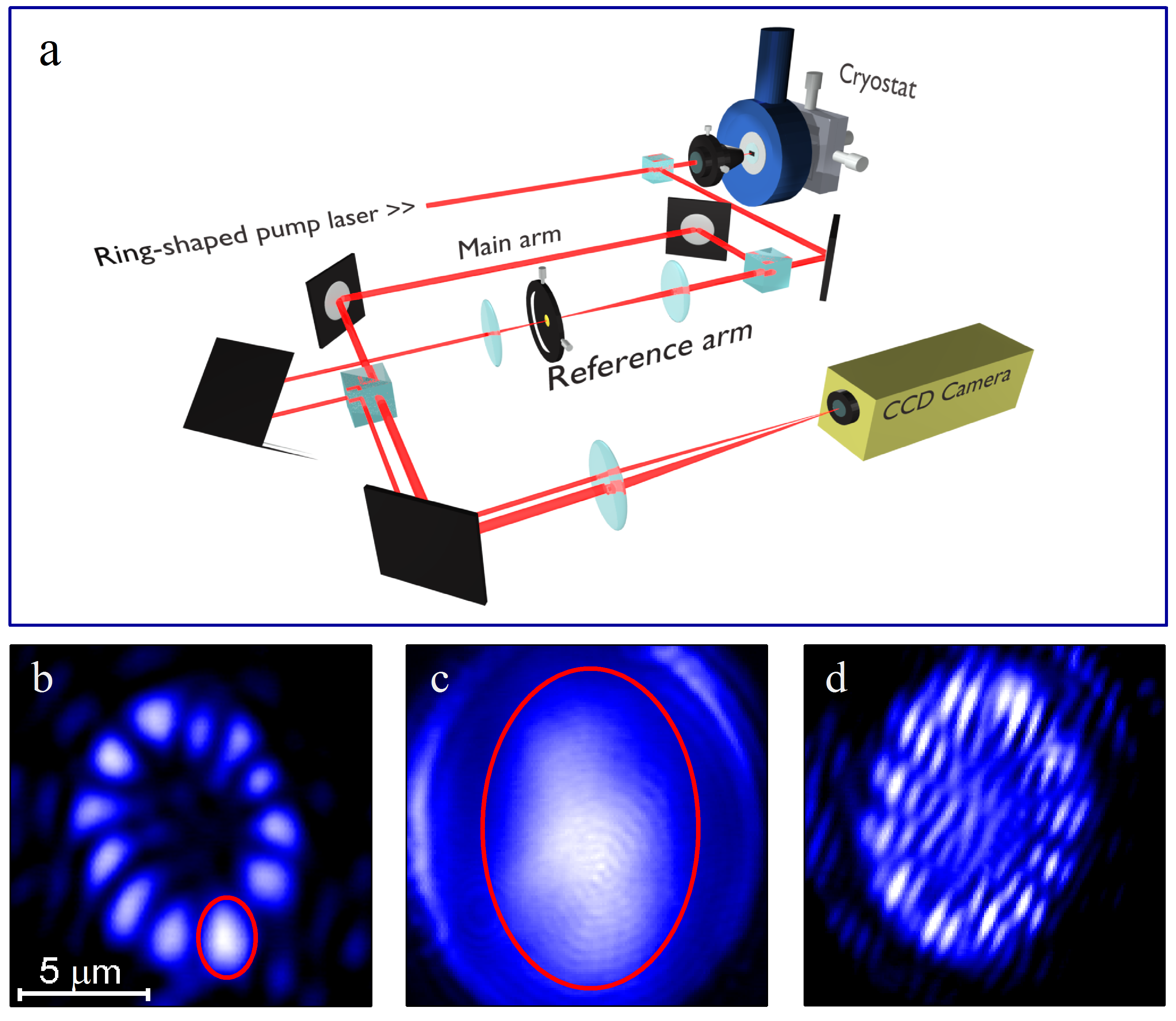}
\caption{(a) Graphical depiction of the experimental setup employed to perform the measurements. The ring-shaped laser, obtained using an axicon lens, is focused on the sample through a 0.5 NA microscope objective. The same objective collects the PL signal, which is then sent towards the detection side. Operating principle of the modified Mach-Zehnder interferometer: the luminescence coming from the sample, shown in (b), is overlapped with a selected region of itself (marked by the red ellipse in (a) and (b)) to obtain the resulting interferogram (c). Note that the real space density in (b) is acquired here for a pump power of 300 $\mu$W, showing that the pattern remains identical for pump powers higher than condensation threshold.}
\label{fig:figure_setup}
\end{figure}

In order to provide a full experimental characterization of the polariton condensed phase, we make use of the experimental setup depicted in Fig.~\ref{fig:figure_setup}a. The sample, the same CdTe semiconductor microcavity of~\cite{kasprzak_condensation_2006}, is kept in a cold-finger cryostat at liquid helium temperature ($\approx$ 4 K). The non-resonant optical excitation is provided by a quasi continuous-wave Ti-Sapphire monomode laser. In order to engineer the ring-shaped excitation beam, we employ an optical loss-less technique which involves the use of an axicon, a conical lens that is able to convert an input Gaussian beam into an output laser beam with a ring intensity distribution. We are able to finely tune the size and ensure the homogeneity of the laser ring by finely adjusting the axicon position with respect to the subsequent focusing lenses using x-y-z micrometric translation stages. The properly shaped laser beam is then focused on the surface of the sample through a high numerical aperture (0.5 NA) microscope objective. The same objective collects the luminescence emitted by the sample that is sent towards the detection side of the setup, whose core is a modified Mach-Zehnder interferometer. Since we are using non-resonant excitation, a traditional homodyne detection scheme is not applicable to our case. Therefore, we implemented a customized version of the Mach-Zehnder interferometer in which one of the two arms, the so-called reference arm (Fig.~\ref{fig:figure_setup}c), is a magnified version of the other one, which we call main arm (Fig.~\ref{fig:figure_setup}b) or PL arm (see sketch of Fig.~\ref{fig:figure_setup}a). The magnification of the reference is set to four times with respect to the main arm by the choice of the focal lengths of the two-lens telescope. A pin-hole allows us to filter in real-space the portion of the magnified PL (giving rise to some residual airy fringes visible in the outer margins in Fig.~\ref{fig:figure_setup}c) that we want to overlap with the main arm to get the interference pattern, an example of which is shown in Fig.~\ref{fig:figure_setup}d. Fine tuning of the output optical cube allows to select at will the overlap conditions between the two arms, while lateral and vertical shifts of the retroreflector result in a controlled wavevector mismatch between the two interfering beams, making it possible to set the density and the direction of the interference fringes.
\begin{figure}[tb]
\includegraphics[width=0.48\textwidth]{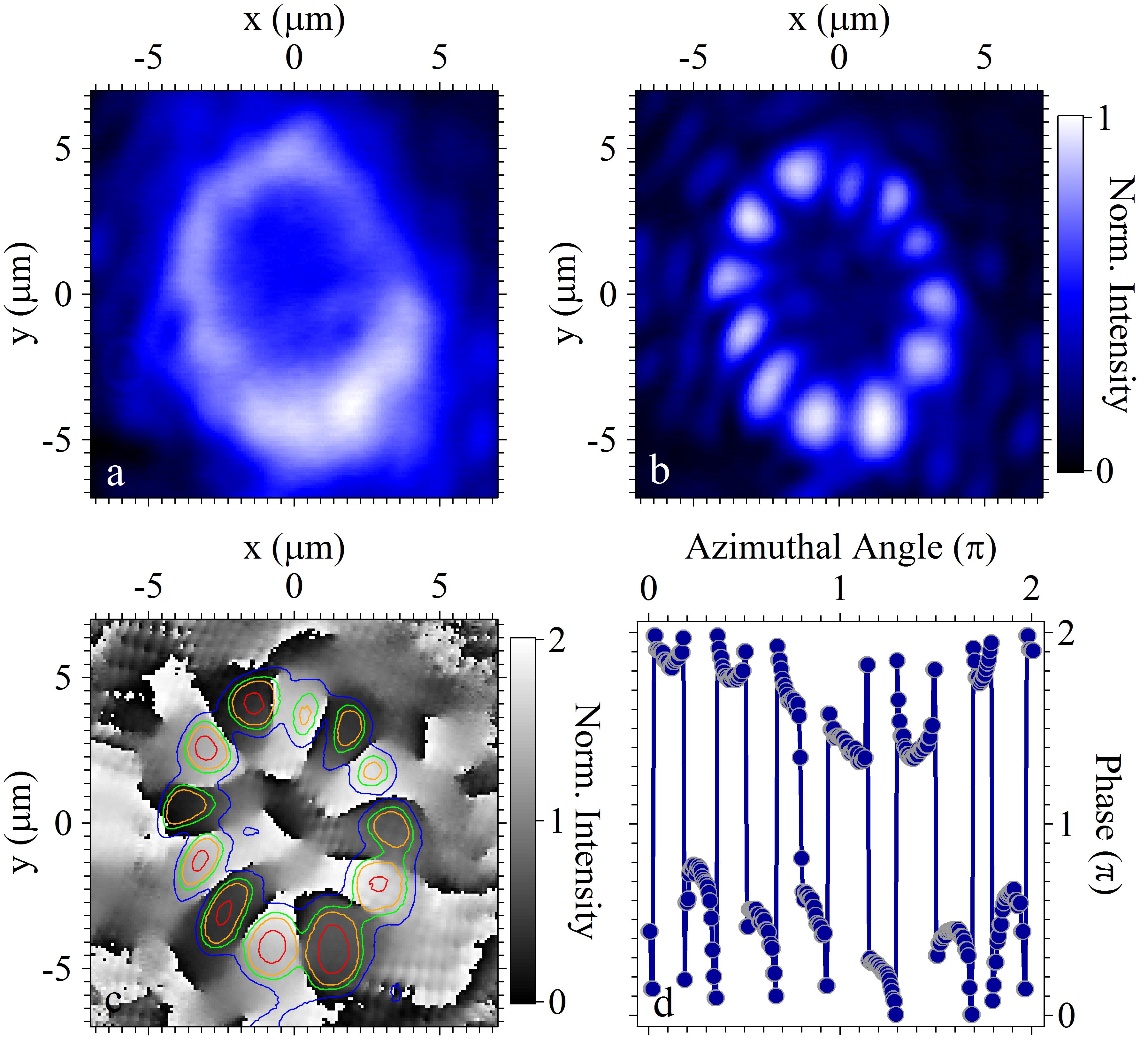}
\caption{(a) Polariton density below condensation threshold under ring-shaped laser excitation. (b) Polariton density above condensation threshold. The spontaneous formation of a spatial pattern, made up of 12 lobes, can be clearly observed. (c) Phase map corresponding to the polariton density shown in (b) where a density contour plot is superimposed in order to allow identification of the density lobes in the phase structure. (d) Phase profile along the patterned density ring, where the phase jumps between one lobe and the following can be identified.}
\label{fig:figure_exp}
\end{figure}

With the aforementioned setup, we are able to determine both the density and the phase of the polariton condensate. The phase extraction from the interferogram image is based on digital off-axis processing techniques, as in our previous works~\cite{lagoudakis_vortices_2008,lagoudakis_halfvortices_2009}. For a pump power of 100 $\mu$W we are below the polariton condensation threshold and the resulting detected photoluminescence (PL) signal is the one shown in Fig.~\ref{fig:figure_exp}a. The PL signal, as the excitation laser, is in its turn ring-shaped. Despite the homogeneity of the prepared laser ring (not shown), it can be seen that the PL is slightly affected by the natural disordered potential present at this specific position in the sample. This results in a higher intensity at one of the two sides of the PL ring. As condensation threshold is reached (as shown in Fig.~\ref{fig:figure_exp}b), for a pump power of around 250 $\mu$W, we observe the sudden spontaneous appearance of a single-energy condensed state of polaritons featuring a clear pattern of 12 lobes along the ring. Another feature worth noting in Fig.~\ref{fig:figure_exp}b is that the polariton density in the lobes keeps being affected by the presence of the natural disorder: the local blueshift along the ring is such to lock the phase between the various lobes and form a single-energy state in agreement with the mode-synchronization mechanism~\cite{baas_synchronized_2008}, as also observed in the case of one-dimensional polariton condensates in a disordered potential~\cite{manni_polariton_2011}.

A tomographic reconstruction of the real-space emission at the energy of the condensate (see supplemental material) shows that, for pump powers just above threshold, polariton condensation indeed occurs in a single-energy state. The energy resolved pattern matches perfectly well the spectrally integrated PL. Further increasing the excitation power we keep observing the same spatial pattern, up to almost two times the condensation threshold. For even higher excitation powers, condensation starts to take place in multiple spatially overlapping modes. Condensation in a single energy mode allows us to directly extract the phase even from an energy-integrated interferogram, since the corresponding phase distribution only comes from the condensed state. In order to do so, we make use of the modified Mach-Zehnder interferometer, where the PL signal is overlapped with one of the lobes of the patterned condensate, providing an approximately flat reference phase for the whole interferogram. In the case shown, the obvious choice is to select the largest lobe of the condensate pattern to overlap with the main arm PL. The interference pattern obtained is then used to retrieve the phase distribution of the condensate, shown in Fig.~\ref{fig:figure_exp}c. In the phase distribution, the corresponding 12-lobe density structure can be easily identified, each lobe having a well-defined phase. Between each lobe and its two nearest neighbors it is expected from theory to have a phase jump of the order of $\pi$, in agreement with the circular phase profile extracted along the patterned ring (see Fig.~\ref{fig:figure_exp}d).

Polariton condensates generated with a non-resonant pump represent a highly non-equilibrium system. Whilst the polariton condensate itself can be accurately described by a mean-field approximation, it is essential in a full theoretical treatment to also account for the incoherent excitations generated by non-resonant pumping that couple to the condensate. An intuitive description makes use of a Gross-Pitaevskii type equation for the condensate mean-field, coupled to a classical rate equation for the dynamics of an exciton reservoir. Such a description has appeared in Ref.~\cite{wouters_spatial_2008} and formally reads as follows
\begin{align}
i\hbar \frac{\partial \psi(\mathbf{r},t)}{\partial t} & =\left[\hat{E}_{LP} + V(\mathbf{r}) + \frac{i\hbar}{2}\left(R_{R}n(\mathbf{r},t) - \gamma_{c}\right)\right] \psi(\mathbf{r},t) \label{eq:schrodinger} \\
\frac{\partial n(\mathbf{r},t)}{\partial t} & = -\left(\gamma_{c} + R_{R}|\psi(\mathbf{r},t)|^{2}\right) n(\mathbf{r},t) + P(\mathbf{r}) \label{eq:reservoir}
\end{align}%
where $\psi(\mathbf{r},t)$ is the mean-field representing the polariton condensate and $n(\mathbf{r},t)$ is the intensity distribution of an incoherent excitonic reservoir. For simplicity we neglect the polarization degree of freedom. $\hat{E}_{LP}$ is the polariton kinetic energy operator, which represents the non-parabolic dispersion of lower branch polaritons. $V(\mathbf{r})$ represents an induced effective potential, given by the mean-field shift caused by polariton interactions ($g$), the interaction of polaritons with the reservoir ($g_{R}$) and additional pump induced shift (G)~\cite{wouters_spatial_2008}; $V(\mathbf{r}) = \hbar g|\psi(\mathbf{r},t)|^{2} + \hbar g_{R}n(\mathbf{r},t) + \hbar GP(\mathbf{r})$, where $g$, $g_{R}$ and $G$ are constants. Although interactions between polaritons inside the condensate are present, the dominant effective potential terms are arising from the effect of the exciton reservoir on condensed polaritons~\cite{roumpos_gain_2010}. $\gamma_{c}$ and $\gamma_{R}$ represent the decay rates of condensed polaritons and reservoir excitons, respectively. $R_{R}$ is the condensation rate.
Equations \ref{eq:schrodinger} and \ref{eq:reservoir} can be solved numerically in time until a steady state is reached starting from an initial random phase (white noise) and low intensity distribution for $\psi(\mathbf{r},0)$.
\begin{figure}[tb]
\includegraphics[width=0.48\textwidth]{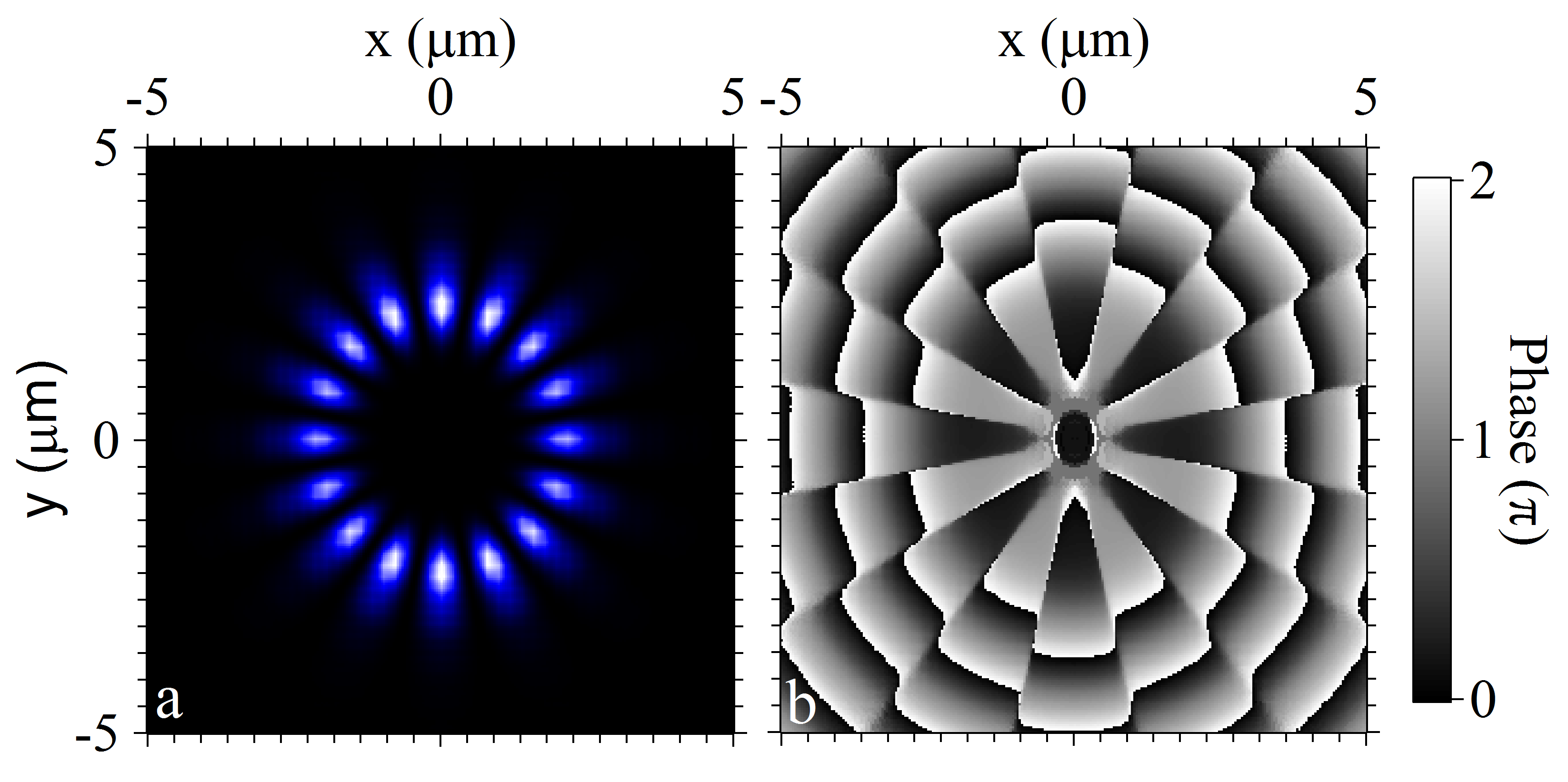}
\caption{Results of the simulations performed with a Gross-Pitaevskii based model under non-resonant (slightly elliptical) ring-shaped laser excitation. (a) Polariton density of a 16-lobe spatial pattern and  (b) phase distribution.}
\label{fig:figure_theory}
\end{figure}
Using a slightly elliptical ring shaped pump and solving the equations for the steady state indeed leads to ring shaped multi-lobed spontaneous polariton patterns for parameters corresponding to our experiment~\cite{simulation_param}. Fig.~\ref{fig:figure_theory} shows the resulting polariton intensity and phase calculated with this method. Within our theory, the number of lobes is quite sensitive to the profile of the pump. For slightly larger sized rings we have computed up to twenty-four lobes. Experimentally, we also find that the number of lobes is sensitive to the position on the sample and the profile of disorder (which we do not include in the theoretical model for simplicity). Theoretically a $\pi$ phase flip is observed between neighboring lobes, in agreement with the experimentally measured phase distribution, shown in Fig.~\ref{fig:figure_exp}c.

Qualitatively, the pattern formation can be interpreted as a consequence of condensation into a single energy state with periodic boundary conditions. The optically induced localization generates a spectrally discrete set of "standing wave" modes, which differ in the number of angular nodes. In a cylindrically symmetric trap, each energy level would be two-fold degenerate, since eigenstates composed from symmetric and antisymmetric linear combinations of states with equal and opposite orbital angular momentum are both possible. Without any mechanism to break this symmetry, one would expect a cylindrically symmetric density distribution to be excited and no pattern formation to occur. However, in the geometry we are dealing with here, the symmetry is not perfectly cylindrical due to the slight elliptical shape of the pump laser beam. In this case the degeneracy between symmetric and antisymmetric states is lifted and polariton condensation takes place into a state featuring a multi-lobe spatial pattern. The single-energy condensed state, as reproduced in the theoretical simulations, results from a complex interplay between the localization effect of the narrow excitation ring profile and the non-equilibrium pump-decay processes for the polariton system.

The situation can be considered, to some extent, analogous to the case of condensation in a polariton microwire~\cite{wertz_spontaneous_2010} where the eigenstates are non-degenerate one-dimensional standing waves. When polariton condensation occurs into one of those modes it is clear that a pattern in intensity appears, given by the standing wave profile. In our case, instead of using a real potential, we are able to achieve a spontaneous pattern formation in a one-dimensional system with periodic boundary conditions using all-optical means, as we are able to provide with the same excitation beam both the pumping and the localization.

In summary, in this work we have reported on the spontaneous pattern formation of an all optically induced quasi-one dimensional polariton condensate with periodic boundary conditions, under non-resonant optical excitation. The slightly elliptical ring-shaped intensity profile of the laser, that is used to pump the system, is also responsible for the symmetry breaking. In these conditions polariton condensation is found to occur in a single-energy state, featuring a multi-lobe density pattern. The theoretical description we provide, based on a mean-field GPE, is able to reproduce the experimental findings, capturing the physics of the observed spontaneous pattern formation phenomenology in a polariton condensate.

We would like to thank Y.~Leger, M.~Wouters and V.~Savona for the fruitful discussions. This work was supported by the Swiss National Science Foundation through NCCR ``Quantum Photonics.''

\bibliography{PatternFormation}

\begin{thebibliography}{21}
\bibitem[{sim()}]{simulation_param}
\bibinfo{note}{We used the following parameters in the simulations:
  $\hbar\gamma_{c} = 1$ meV, $\hbar\gamma_{R} = 10$ meV, $\hbar R_{R} = 0.1$
  meV$\mu m^{2}$, $G = 0.005 \,\mu m^{2}$, $\hbar g_{R} = 0.022$ meV$\mu
  m^{2}$. The polariton dispersion was calculated with the standard two
  oscillator model with effective cavity photon mass $3\times10^{-5}$ of the
  free electron mass and 26 meV Rabi splitting.}
\end{thebibliography}

\newpage
\section{Supplemental material}
\begin{figure}[b]
\includegraphics[width=0.48\textwidth]{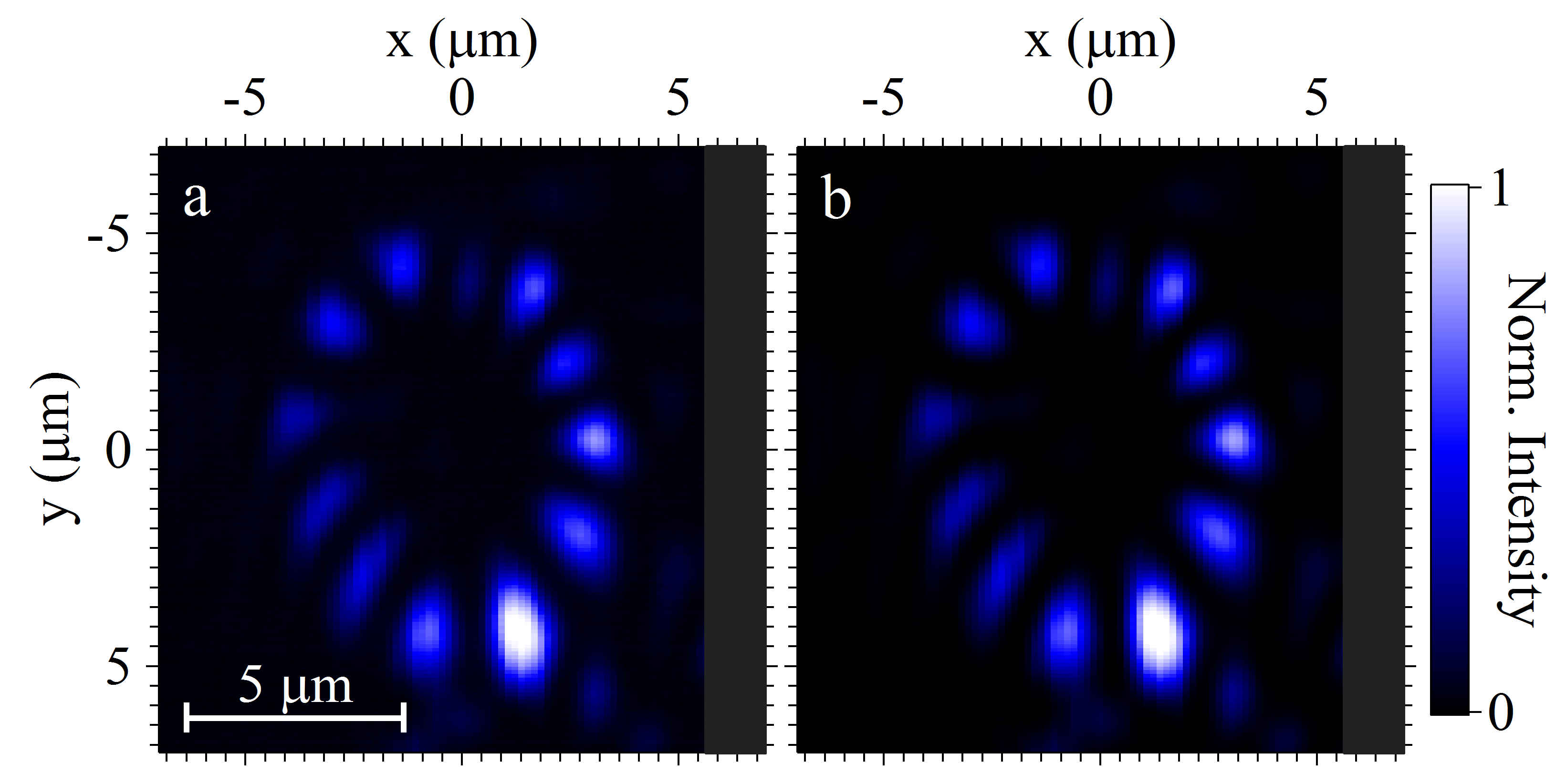}
\caption{(a) Spectrally integrated polariton density. (b) Spectrally resolved polariton density at the energy of the condensate. Both images are acquired above condensation threshold, for a pump power of 250 $\mu$W. There is an evident matching between the two images. Note: the grey shaded region has been added to the graphs to keep the same axis formatting as in Fig.~\ref{fig:figure_exp} and it corresponds to missing data, anyway outside the region of interest.}
\label{fig:figure_suppl}
\end{figure}

Tomographic reconstruction of the real-space condensate density
In order to verify that the spontaneous pattern formation occurs in a single-energy state and that it constitutes the only significant contribution to the spectrally integrated PL, for pump powers close to condensation threshold, we performed a polariton density tomographic reconstruction with energy resolution. The procedure consists in collecting the PL signal of the main arm and focusing it onto the slits of the spectrometer (oriented in the y direction of Fig.~\ref{fig:figure_suppl}) in order to reproduce the real-space polariton density on them. By scanning the position of the focusing lens (along the x direction of Fig.~\ref{fig:figure_suppl}), different real-space slices enter the slits of the spectrometer, allowing to retrieve the energy information for each real-space slice. The results obtained, for a pump power of 250 $\mu$W, are summarized in Fig.~\ref{fig:figure_suppl}. In Fig.~\ref{fig:figure_suppl}a we show the spectrally integrated real-space image, obtained summing over all the energy range acquired. In Fig.~\ref{fig:figure_suppl}b, instead, we selected only the real-space emission at the energy of the patterned condensate. It can be seen, by comparing the two figures and also referring to the spectrally integrated CCD image of Fig.~\ref{fig:figure_exp}b, that the spontaneous 12-lobe pattern really appears in a single-energy condensed state. An additional direct proof of the coherence between all the lobes of the pattern comes from the interference fringes that appear all over the PL signal, when overlapping it with a magnified version of one of its lobes. The fact that the main contribution to the PL signal comes from a single state, allows us to perform the phase extraction from a spectrally integrated interferogram, like the one shown in Fig.~\ref{fig:figure_setup}d.

\end{document}